\documentclass[a4paper]{article}

\usepackage{graphicx}
\usepackage{amsmath}
\usepackage{mathrsfs}
\usepackage{booktabs}
\usepackage{CJKutf8}
\usepackage{pdfpages}
\usepackage{natbib}
\usepackage{hyperref}
\usepackage{enotez}
\let\footnote=\endnote

\usepackage{tabularx}
\usepackage{array}
\usepackage{soul}
\usepackage{xcolor}
\usepackage{tcolorbox}
\usepackage{makecell}

\usepackage{geometry}
\usepackage[capitalize]{cleveref}
\crefname{section}{Sec.}{Secs.}
\Crefname{section}{Section}{Sections}
\Crefname{table}{Table}{Tables}
\crefname{table}{Tab.}{Tabs.}

\author{
\small
Abeba Birhane,
Mozilla Foundation \& \\\small
 School of Computer Science and Statistics,\\\small
Trinity College Dublin, Ireland\\\small
birhanea@tcd.ie
\and\small
Jelle van Dijk\\\small
Faculty of Engineering Technology,
\\\small
University of Twente, Enschede, Netherlands\\\small
jelle.vandijk@utwente.nl
\and\small
Frank Pasquale\\\small
Cornell Tech and Cornell Law School\\\small
New York, USA,\\\small
frank.pasquale@gmail.com
}
\date{}

\begin{document}
\title{Debunking Robot Rights Metaphysically, Ethically, and Legally}

\maketitle

\begin{abstract}
In this work we challenge arguments for robot rights on metaphysical, ethical and legal grounds. \textit{Metaphysically}, we argue that machines are not the kinds of things that may be denied or granted rights. Building on theories of phenomenology and post-Cartesian approaches to cognitive science, we ground our position in the lived reality of actual humans in an increasingly ubiquitously connected, controlled, digitized, and surveilled society.  \textit{Ethically}, we argue that, given machines’ current and potential harms to the most marginalized in society, limits on (rather  than rights for) machines should be at the centre of current AI ethics debate. From a \textit{legal} perspective, the best analogy to robot rights is not human rights but corporate rights, a highly controversial concept whose most important effect has been the undermining of worker, consumer, and voter rights by advancing the power of capital to exercise outsized influence  on politics and law. The idea of robot rights, we conclude, acts as a smoke screen, allowing theorists and futurists to fantasize about benevolently sentient machines with unalterable needs and desires protected by law. While such fantasies have motivated fascinating fiction and art, once they influence legal theory and practice articulating the scope of rights claims, they threaten to immunize from legal accountability the current AI and robotics that  is fuelling surveillance capitalism, accelerating environmental destruction, and entrenching injustice and human suffering.

\end{abstract}



\section{Introduction}
\label{sect:intro}

For some theorists of technology, the question of AI ethics must extend beyond consideration of what robots and computers may or may not do to persons.   Rather,  persons may have some ethical and legal duties to robots–up to and including recognizing robots’ “rights”. David Gunkel hailed this Copernican shift in robot ethics debates as “the other question: can and should robots have rights?”~\citep{gunkel2018other}. Or, to adapt John F. Kennedy’s classic challenge
: Ask not what robots can do for you, but rather, what you can, should, or must do for robots.

In this work we challenge arguments for robot rights on metaphysical, ethical and legal grounds. Metaphysically, we argue that machines are not the kinds of things that could be denied or granted rights. Ethically, we argue that, given machines’ current and potential harms to the most marginalized in society, limits on (rather than rights for) machines should be at the centre of current AI ethics debate. From a legal perspective, the best analogy to robot rights is not human rights but corporate rights, rights which have frequently undermined the democratic electoral process, as well as workers’ and consumers’ rights. The idea of robot rights, we conclude, acts as a \textit{smoke screen}, allowing theorists  to fantasize about benevolently sentient machines, while so much of current AI and robotics is fuelling surveillance capitalism, accelerating environmental destruction, and entrenching injustice and human suffering.

Building on theories of phenomenology and post-Cartesian approaches to cognitive science, we ground our position        in the lived reality of actual humans in an increasingly ubiquitously connected, automated and surveilled society. What we find is the seamless integration of machinic systems into daily lives in the name of convenience and efficiency. The  last thing these systems need is legally enforceable “rights” to ensure persons defer to them. Rights are exceptionally powerful legal constructs that, when improvidently granted, short-circuit exactly the type of democratic debate and empirical research on the relative priority of claims to autonomy that are necessary in our increasingly technologized world~\cite{dworkin1984rights}. Conversely, the ‘fully autonomous intelligent machine’ is, for the foreseeable future, a sci-fi fantasy, primarily functioning now as a meme masking the environmental costs and human labour which form the backbone  of contemporary AI. The robot rights debate further mystifies and obscures these problems.  And it paves the way for a normative rationale for permitting powerful entities developing and selling AI to be \textit{absolved} from accountability and responsibility, once they can program their technology to claim rights to be left alone by the state.

Existing robotic and AI systems (from large language models (LLMs), “general-purpose AI”, chatbots, to humanoid robots) are often portrayed as fully autonomous systems, which is part of the appeal for granting them rights. However, these systems are never fully autonomous, but always \textit{human-machine systems} that run on exploited human labour and environmental resources. They are socio-technical systems, human through and through – from the source of training data,  to model development, societal uptake following deployment, they necessarily depend on humans. Yet, the “rights” debate too often proceeds from the assumption that the entity in question is somewhat autonomous, or worse that it  is devoid of exploited human labour and not a tool that disproportionately harms society’s disenfranchised and minoritized. Current realities require instead a reimagining of technologies from the perspectives, needs, and rights of the most marginalized and underserved. This means that any robot rights discussion that overlooks underpaid and exploited populations that serve as the backbone for “robots” (as well as the environmental cost required to develop AI) risks being disingenuous. As a matter of public policy, the question should not be whether robotic systems deserve rights, but rather, if we grant  or deny rights to a robotic system, what consequences and implications arise for people owning, using, profiting, developing, and affected by actual robotic and AI systems?

The time has come to change the narrative, from “robot rights” to the duties and responsibilities of the corporations and powerful persons now profiting from sociotechnical systems (including, but not limited to, robots). Damages, harm and suffering have been repeatedly documented as a result of the integration of AI systems into the social world. Rather than speculating about the desert of hypothetical machines, the far more urgent conversation concerns robots and AI as concrete artifacts built by powerful corporations, further invading our private, public, and political space, and perpetuating injustice. A purely intellectual and theoretical debate obscures the real threat: that many of the actual robotic and AI systems that powerful corporations are building are harming people both directly and indirectly, and that a premature and speculative robot rights discourse risks even further unravelling our frail systems  of accountability for technological harm.

The rise of “gun owners’ rights” in the US is but one of many prefigurative phenomena that should lead to deep and abiding caution about fetishization of technology via rights claims. U.S. gun owners’ emotional attachments to their weapons are often intense (Franks, 2019). The US has more gun violence than any  other developed country, and endures frequent and bloody mass shootings. Nevertheless, the US Supreme Court has advanced a strained interpretation of the US Constitution’s Second Amendment to promote gun owners’ rights above public safety. We should be very careful about robot rights discourse, lest similar developments empower judiciaries to immunize exploitive, harmful, and otherwise socially destructive technologies from necessary regulations and accountability.

The rest of the paper is structured as follows: Section~\ref{sect:robot_rights} sets the scene by outlining the robot rights debate. In Section~\ref{sect:robot_at_issue}, we delve into what exactly “robot” entails in this debate. Section~\ref{sect:machines_like_us} clarifies some of the core underlying errors committed by robot rights advocates surrounding persons, human cognition, and machines and presents our arguments, followed by a brief overview of embodied and enactive perspectives on cognition in Section~\ref{sect:embodied}. In Section~\ref{sect:posthumanism}, we examine posthumanism, a core philosophical approach by proponents of robot rights and illustrate the problems surrounding it. Section~\ref{sect:legal} outlines the legal arguments, which is followed by Section~\ref{sect:implications} which illustrates how rights talk “cashes out” in actionable claims in courts of law, demonstrating the real danger of robot rights talk. We conclude in Section~\ref{sect:conclusion} by emphasizing the enduring irresponsibility of robot/AI rights talk.

\section{Robot Rights: the Debate}
\label{sect:robot_rights}

For many outside the small circle of robot rights discourse, the very idea of debating whether (future) artificially intelligent entities should be granted or denied rights might seem bizarre, fanciful, or even obscene. Nevertheless, within the narrow corners of AI ethics/machine ethics/robot ethics, the notion of robot rights has sparked ferocious debates~\endnote{In our broad definition, robots include (but is not limited to) large language models, humanoid robots, “AI assistants”, household gadgets such as Siri, Nest, Roomba, drones (such as those used in warfare), social bots,  chatbots, algorithms (such as those central to online platforms), generative AI, “general purpose AI”, and what has recently been dubbed as “frontier AI”.  For a detailed discussion of the definitional issues raised by the term robot, see Richards \& Smart, ”How Should the Law Think About Robots,” in \textit{Robot Law}~\cite{calo2016robot}}.

The idea of robot personhood, and closely aligned ideas of robot rights, has been a long-standing, if fringe, position. It was directly advocated by~\citet{inayatullah2001rights}, and discussed in the early 1990s by legal scholar~\citet{solum1991legal}. Numerous scholars have discussed the possibility and propriety of granting robots some form of legal rights~\cite{gunkel2018other}. Arguments for robot rights tend to come in two broad forms. One set of arguments is built around the idea that humans will be more virtuous and respectful to one another if they are required to respect certain machines (e.g., those that use sensors to gather data about their environment, process that information, and execute certain actions in a given environment). This set of arguments, focused on how granting rights to inanimate objects may cultivate certain dispositions, skills,  and capacities in human beings, was directly addressed in earlier work by~\citet{birhane2020robot}.

Another set of arguments focuses on robots themselves, rather than their effects on persons. Some scholars argue  that there is a mix of qualities of present and/or future robots that could merit the type of respect that legal systems usually offer in the form of rights to do certain actions, or to prevent others from doing certain actions~\cite{bennett2020recognising,turner2018robot}. This and similar sets of arguments constitute the main concern of this paper. To address it properly, we first carefully examine what “robots” entail in the next section.

\section{The Robots at Issue}
\label{sect:robot_at_issue}

Both the artificially intelligent entity in question and the reasoning for granting/denying rights varies.  Clarifying  the “robot” in question is critical.  Actually existing robots include machines constructing cars on assembly lines; robots meant to simulate animals (for example, “robot dogs” manufactured by companies like Boston Dynamics  (for war and business) and once manufactured by Sony (the Aibo, for home entertainment); humanoid robots (like the discontinued Pepper); service delivery robots (such as Starship and other drones); and a catch-all category of machines with sensors, information processors, and actuators and massive machine learning systems such as LLMs. 

As a practical and immediate matter, the question of granting rights to current machines (not the supposedly   conscious and super intelligent futuristic type but the ubiquitously integrated algorithmic systems) signals a grave danger.  To grant these machines rights seems to hinge on an argument of complexity:  in large-scale AI systems it is no longer clear whether the humans in charge of these systems can still be held responsible for the deeds of their ‘autonomous’ systems. Consider, for example, a set of police robots embedded with latest a version of GPT$-\{n\}$, which might ‘act’ in unpredictable ways once it interacts with persons. If a particularly bad outcome  occurs, the manufacturers and deployers of the robots might plead that the robots’ actions were unpredictable. This serves as a way of evading responsibility leaving nobody accountable for harmful ’actions’.

Such a strategy deflects responsibility from the human owners, creators, operators, vendors, and/or beneficiaries of the machine.  Moreover, what does it mean to grant rights to generative models that produce conspiracy theories, lies, and social stereotypes~\cite{bender2021dangers,weidinger2022taxonomy};  systems that mass harvest data without awareness or consent of the data subject and produce models that perpetuate harmful stereotypes~\cite{birhane2021large,hundt2022robots}; physical robots that could take on all the roles of drones and other technology  that now hyper-surveil marginalized communities~\cite{arnett2020race,ajunwa2017limitless}; algorithmic systems that threaten privacy~\cite{kalluri2023surveillance,veliz2020privacy,southerland2020intersection}; bots that spread disinformation~\cite{calo2020robots,howard2020lie}; and generally speaking AI systems that are created at a massive environmental and labour cost~\cite{dodge2022measuring,lacoste2019quantifying,crawford2021atlas,gray2019ghost} all of which disproportionately impact individuals and communities at the margins of society?

A somewhat different form of advocacy for robot rights is rooted in a version of anticipatory philosophizing and futurist premises.  The argument goes:  science and technology are advancing rapidly enough to ensure that, within our lifetimes, some machines will so convincingly emulate human action, expression, and interests that human beings will (and should) feel obliged to respect these machines to some extent. If robots exhibit a set of humanlike qualities, then there is no reason to deny  robots rights.

Assuming that the potential beneficiaries of “robot rights” will be far more advanced machines than current ones is the key argumentative move that makes robot rights conceivable at all. However, that concession may well itself obscure more than it reveals, since it presumes a level of autonomy and distinctness of the machine that has been absent in the history of human-computer interaction, which is likely to be so far into the future (if it is possible at  all). That is to say, machines are \textit{never} fully autonomous but always human-machine systems that rely on human power and resources in order to function. Automation and the idea of automata, from its early conception, relied  on a clever trick that erased the labourers toiling away in the background, the people performing crucial tasks for  a machine to operate~\cite{jones2020ghost,sadowski2018potemkin,taylor2018automation}. Surveying the historical genealogy of mechanical calculations,~\citet{daston2017calculation} emphasizes that far from relieving the mental burden, automation, shifted the  burden to other shoulders (often women who were underpaid), maintaining the ghost in the machine. Furthermore, as demonstrated by Bainbridge’s seminal work on aircraft automation, “the more advanced a control system is, so the more crucial may be the contribution of the human operator”~\cite{bainbridge1983ironies}. Bainbridge’s concept of the \textit{ironies of automation}, gets at the heart of intentionally hidden human labour in an appeal to portray machines as autonomous. In reality as~\citet{baxter2012ironies} building on Bainbridge’s classic work make explicit,  “[T]he more we depend on technology and push it to its limits, the more we need highly-skilled, well-trained, well- practised people to make systems resilient, acting as the last line of defence against the failures that will inevitably occur.”~\cite{baxter2012ironies}. 40 years later, Bainbridge’s point remains. Current state-of-the-art AI systems such as ChatGPT operate on the backbone of exploited and underpaid labour often outsourced to low-to-middle income countries such as Kenya ~\cite{perrigo2023exclusive}. Effective automation of control processes necessarily requires humans at various steps including developing, maintaining and stepping in  when ‘autonomous’ systems inevitably experience failure~\cite{hancke2020ironies,baxter2012ironies}. Today’s ‘autonomous’ systems such as ‘autonomous vehicles’ still depend on human intervention and feedback, ranging from “safety drivers” to volunteer reporters, reinforcement learning from human feedback (RLHF) (the case also for most publicly released large language models),  and the need for human input is unlikely to disappear entirely although it might change form~\cite{tubaro2019micro}. To fixate on rights claims by a particular machine in this human-machine-society assemblage, obscures the more ethically urgent questions of who is doing the crucial data work, whether their compensation and conditions  of work are fair and safe for them, and whether the socio-technical configuration as a whole sustainably contributes to human well-being.

Not only does seemingly autonomous AI rely on high-paid, high-skilled engineers and scientists, it also relies in crucial ways on underpaid, undervalued, and less-visible labour which go by various names including \textit{ghost work}, \textit{microwork}, and \textit{crowd-work}~\cite{irani2015cultural,gray2019ghost,perrigo2023exclusive}. From labeling images, identifying objects in images, annotating data, to content moderation, such human labour “help AI get past those tasks and activities that it cannot solve effectively and/or efficiently”~\cite{tubaro2019micro}, step in to fill in when AI fails ~\cite{irani2015cultural} and are automation’s \textit{‘last mile’}~\cite{gray2019ghost}. Such work constitutes the backbone of current AI – without which, AI would cease to function. Yet, from \textit{Amazon’s MTurk}, to \textit{Clickworker},  \textit{AppJobber},  to \textit{CrowdTap},  such labour is under-compensated and often goes unrecognized in the AI pipeline. People doing such work, in most cases are not formally considered as formal employees but independent contractors further adding to their precarious working and living conditions. Indeed, close inspection of the actual material foundations of AI reveals that advocacy for robot rights is much closer to advocacy for corporate or capital rights to subordinate the most exploited, than it is to labor or human rights, the latter of which are actually at peril more than ever, in todays’ capitalist and corporate world driven by profit maximization. 

Furthermore, putting the exploitative structure of microwork aside, training data, foundation to  AI systems, is often sourced in questionable manners; uncompensated and unregulated. AI models built on such data necessarily inherit dataset problems such as encoding and exacerbating societal and historical stereotypes (with downstream impacts that negatively and disproportionately disadvantage marginalized communities). The deep learning revolution that transformed image detection, identification, recognition, and generation for example, is only possible through data theft ~\cite{paullada2021data} through continued mass scraping of user uploaded images, all sourced and used to build large AI systems without consent or awareness of image owners~\cite{birhane2021large,birhane2023into}. Of some of the major large scale datasets currently used  to train and validate computer vision models, none was sourced consensually as of 2021, as detailed in Table~\ref{tab:concrete_summary}) (see~\citet{birhane2021large} for more).

\begin{table} [htbp]
\centering
\setlength{\tabcolsep}{3pt}
\caption{Large scale datasets containing people’s images.}
\label{tab:concrete_summary}
\scriptsize
\begin{tabular}{@{}lrrr@{}}
\toprule
\textbf{Dataset} & \makecell{\textbf{Number of images}\\ \textbf{(in millions)}} & \makecell{\textbf{Number of categories}\\ \textbf{(in thousands)}} & \makecell{\textbf{Number of}\\ \textbf{consensual images}}\\ 
\midrule
\textbf{JFT-300M} & 300+           & 18          & 0 \\ 
\textbf{Open Images} & 9           & 20           & 0 \\ 
\textbf{Tiny-Images}\protect\footnotemark & 79           & 76          & 0 \\ 
\textbf{Tencent-ML} & 18            & 11           & 0 \\ 
\textbf{ImageNet-(21k,11k,1k)}  & (14,12, 1)           & (22,11,1)          & 0 \\
\textbf{Places} & 11            &  0.4         & 0\\
\bottomrule
\end{tabular}

(Table taken from~\cite{birhane2021large})
\end{table}

Despite all these well-documented problems of present and foreseeable AI research programs, some advocates of robot rights posit that a societal recognition of the powers and value of machines would be advisable on purely human-welfare-enhancing grounds. A person forbidden to kick a robot that looks like a dog, we are assured, is subtly inculcated with a moral obligation to respect anything others might care about, lest it be silicon- or carbon-based.  But the moral lesson could be quite the opposite:  that those in power are so devoted to certain inanimate objects that they have afforded them a status beyond that which is enjoyed by many persons ill-protected by existing legal structures.

Indeed, it is a cruel parody of justice to presume that Saudi Arabia’s potential grant to the robot “Sophia”~\cite{reynolds2018agony} a license to drive, is in some way prefigurative or promoting some later action to grant that kingdom’s women more robust rights. Nor would “rights” for a cleaning robot in the U.S. do anything to help the hundreds of thousands of janitorial workers now suffering in exploitative working conditions. Indeed, some of them are likely already, unwittingly or in  an uncompensated way, providing training data to such robots. Giving a robot a “right to observe” the persons’ it is being programmed to replace is no blow for social justice; it is in fact its opposite, short-circuiting debates about data control and compensation that should be robust~\cite{pasquale2014replace}.

Consequently, an intellectual school and movement so reliant on the rhetoric of emancipation, as the robot rights community is, ends up centering and elevating the power and privileges of capital and corporations. Yet this is not a surprising outcome. Multinational corporations (MNCs) have mastered the “code of capital”~\cite{pistor2019code} to  advance their interests in multiple venues. Investor-State Dispute Settlement (ISDS) provisions in Bilateral Invest ment Treaties (BITs), for example, can guarantee MNCs returns in ways that unravel much-needed environmental,   occupational safety, and health regulation, all based on the MNC’s “right” to enjoy expected returns from their investment. The U.S. Citizens United v. FEC~\citeyear{justia_2010} decision gave corporations near \textit{carte blanche} opportunities to  promote their views and preferred candidates in elections, spawning even more self-aggrandizing judicial interventions  to freeze into place existing power inequalities ~\cite{pasquale2016first}. Similarly, the rhetoric of “robot rights” is another  means to skillfully advance capital’s interests. Indeed, since robots are often capital themselves, “robot rights” all  too often amount to the type of “capital rights” only dreamed of by ambitious libertarians.

\section{The Machines Like us Argument: Mistaking the Map for the Territory}
\label{sect:machines_like_us}
Having addressed definitional issues and introduced relevant scholarly debates, let us then directly meet the claim that future sentient robots should have rights. The argument is that if we can build a human-like machine in principle, there is no reason not to grant such a machine rights. Underlying this argument is the envisionment of a future where machines might be sentient, feel pain, and have desires and wants in ways comparable to how we currently understand ourselves and others as human beings. If we acknowledge basic rights for sentient, feeling, fellow human beings, what right would we have to deny such rights to other, non-human, sentient feeling beings? Such a denial would be unethical, a case of ‘speciecism’. 

Such lines of arguments are advanced by philosophers, scientists and technologists alike. Importantly, the argument builds on making analogies, or even equating, the nature of humans with the properties of intelligent machines, such that it no longer would make sense to draw an ethical distinction between them. Richard Dawkins, for example, justifies the need to consider the moral status of artificial intelligence as: “[T]here is nothing in our brain [...] that could not in principle be reproduced in technology. [...] I see no reason why in the future we shouldn't reach the point where a human-made robot is capable of consciousness and of feeling pain.” Dawkins ~\cite[starting from 0:24]{dawkins20}. 

Although the argument seems reasonable enough, we think it is seriously flawed in a number of ways, where the premise of the argument (there will be sentient machines) builds on a numerous questionable assumptions. First of all, the idea that, since brains are physical systems, therefore it should be possible to recreate them, is problematic in that it  borrows time, resources and technological inventions from the future, that are not at all guaranteed. In fact, we may argue that, given the complexity of the brain, it may take as long as the evolution of the human species (or even longer) to artificially recreate it. Or in any case, there is no indication that it will be sooner. This makes the in principle argument, even if valid, much less meaningful for any reasonable realistic future scenario. Discussing the rights of creatures that will take another few billion years to come into being, seems premature, at the least. There is also no reason in principle why the entire earth could not be reproduced using elements gathered from space; nevertheless, few if any serious thinkers advocate that we discuss the legal or ethical implications of such a second Earth, given the eons it would take to build it. Yet somehow the prospect of copying our own brain seems so intuitively attractive that immense research funding has been invested  in the attempt. Such massive investments on that kind of project, is an ethical problem in and of itself. However, the failures of the Human Brain Project already caution against simplistic assumptions about future replication of humans \textit{in silico}~\cite{yong2019human}. 

Ray Kurzweil poses a seemingly more reasonable challenge than Dawkins. He argues:
\begin{quote}
  If an AI can convince us that it is at human levels in its responses, and if we are convinced that it is experiencing the subjective states that it claims, then we will accept that it is capable of experiencing suffering and joy. At that point AIs will demand rights, and because of our ability to empathize, we will be inclined to grant them.
  \null\hfill(~\citet{kurzweil2015robots})  
\end{quote}

Just as Turing originally used this argument for granting an entity the status of ‘intelligent’, it states that we should grant an entity rights that we would also grant our fellow human beings, on the basis of the idea that we have granted fellow human beings these rights because they show the right sort of behavior that makes them deserve these rights. Many of the criticisms on Turing’s original argument would apply here as well. More importantly, we note that human rights have not been granted to humans because they can prove they are human. So even if Kurzweil’s argument would work for robots, it would at the least create a situation in which human beings get their rights on different grounds than robots would, which in and of itself could be seen as ethically problematic. And indeed we believe it hints at the fact that the argument is just not the right sort of argument for granting rights. Humans are given rights because they would otherwise become oppressed by other human beings. That all the actors in this consideration are all human, is a given, not part of the debate. For any group of humans that in the past has been denied rights (women, enslaved people), it would be problematic that they would need to meet a behavioral requirement in order to ‘join the club’. Groups of human beings who have at some point been granted rights are granted these because they were human all along, and the fact that they were not recognized as such was the problem that needed to be solved. For robots, the question is not whether they will at some point be \textit{discovered} to be human, they would have to be \textit{designed} in such a way so as to meet some criterion of eligibility for rights. 

In the end, instead of discussing human oppression and how to solve it, Kurtzweil’s argument moves the  entire discussion into this abstract philosophical space where “speculating about what a robot really is” becomes a valid approach for analysing the central issues at stake. Meanwhile, at  the practical everyday level of human interaction, we know there is simply no ethical problem of this kind concerning robots. Robots are not the kind of beings who can be oppressed, or denied rights, because other human beings are not acknowledging their humanity. If that were the case, robots would not even be eligible to have rights as robots, they would be discovered to be ‘actually humans all along’. All of this makes little sense, given that, in the real world (not in the abstract, future envisionments), robots are not the kinds of beings that we are: they are human-constructed pieces of equipment. 

What happens in both the mechanical and behavioral arguments is that theorists first abstract away from real-world practices in order to then be able to equate  human phenomena of intelligence, cognition, and emotion and social relating to certain kinds of behaviors of machines. This theoretical move is core of traditional cognitive science, which rests on the idea that in order to achieve scientific understanding of human intelligence, we can model the phenomenon of interest in a machine now, there is nothing wrong in principle, by using a machine model in order to achieve an understanding of a natural phenomenon. But the danger is then to think that the essence of the model is in fact a real aspect of the original phenomenon. 

Such move mistakes the map for the territory. Throughout history we have compared ourselves metaphorically with the most advanced technology   of the time, not just as an epistemological tool, but also making the mistake to insert the mechanics of the technology into ontological claims. The mid-17th century Danish anatomist Nicolas Steno saw the brain as a collection of cavities through which animal spirits flow. A century later when electricity was in vogue, the brain was seen like a galvanic battery. (Again, to say that it helps to understand the brain by thinking about the analogy of a galvanic battery could be helpful for the scientist’s learning process. But to claim that, therefore, in some abstract essential way, the brain is a galvanic battery, would be mixing up map for territory). And by the mid-19th century, nerves were compared to telegraph wires Cobb\cite{cobb2020idea}. After mid-20th century advances in computing, we have come to think of the brain as a computer, an information processing machine. By making this move, the human mind is presented much simpler than it actually is, and the computer 
is attributed more content than it actually deserves. Or as~\citet{baria2021brain} put it:  “the human mind is afforded less complexity than is owed, and the computer is afforded more wisdom than is due.”

Metaphors are pervasive and we don’t seem to do away with them generally. Metaphors illuminate understanding  one kind of thing in terms of another~\cite{lakoff2008metaphors}. But metaphors can also do more harm than good when their exponents forget they are, in fact, only metaphors. Oftentimes, we do forget, sometimes intentionally when such ignorance serves to help advance a larger ideological agenda. While it sometimes makes sense for scientists to use natural phenomena as guiding metaphors for designing technologies, it does not follow that the resulting technologies therefore constitute valid scientific models of the original natural phenomenon. Technologies are tools, and AI technologies are for scientists just that: tools to think with~\cite{van2023reclaiming}. 

Mistaking the map for the territory is one of the most persistent problems in AI. Limited understanding of the complex nature of intelligence as well as over-optimistic and overconfident predictions about AI, as old as the field itself, are undermined by various fallacies and are obstacles to actual progress in AI~\cite{mitchell2019artificial}. Among these fallacies are ‘intelligence is all in the brain’ (which we come back to below) and the fallacy of what~\citet{mitchell2019artificial} calls \textit{wishful mnemonics}, as well as careless use of words towards AI and robots, including attribution of ”learning,” ”seeing,” ”feeling,” and other human qualities based on a misguided apophenia~\cite{mitchell2019artificial}.

For the ethical question however, there is something more nuanced at stake. If computer models would be used, as they should as knowledge vehicles~\cite{van2023reclaiming}, that is, tools, to understand the brain, there would be no ethical problem: we would never consider whether our scientific methods \textit{themselves} should be ‘granted rights’. They are tools of our mind, not minds of their own. But if we make the mistake of conceptually (in the ontological sense) \textit{equating} ourselves with the machines that we use to understand ourselves, for example on the basis that the brain is ‘in the end, also a physical system, just like the computer’, then we can suddenly see the ethical problem arising: because we have already first defined our own selves to be ‘a machine of some kind’, we now suddenly have to discuss whether or not to grant ‘other, non-human machines’ rights, as we grant them to ourselves. 

In other words, to conceive of robots as ‘human-like machines’ (and therefore consider their rights), implicitly means to first perceive human being in machinic terms: complicated biological information processing machines. Once we see ourselves as machines, it becomes intuitive to see machines as ‘in some sense like us’. This, we argue, comes out of the fallacy of mistaking the map for the territory. Just because one is not allowed to walk on the grass in the park, does not mean we now have to consider whether we are allowed to walk on the grass on the map, even if that map is implemented as a physical use object inside the park. We may decide you are not allowed to walk on the map, but the grounds for that decision have nothing to do what we think is appropriate for actual grass. Analogously, we can talk about whether one is allowed to damage real robots (perhaps on grounds of not wasting value, or simply because it is someone else’s property), but such considerations are not similar to discussions about whether one is allowed to ‘damage’ other people or not. We should not mistake the map with the territory, even acknowledging that maps are also concrete objects existing in the territory. Clarity in language is a major step toward defusing the “AI hype”~\cite{raji2022fallacy,mitchell2019artificial}. ~\cite{tucker2022artifice}, for example, suggests (see Table~\ref{tab:intervention}) substituting misleading terms with accurate language brings the focus back to the way actual AI is used in practice.

\begin{table}[ht]
\centering
\caption{Proposed  intervention for AI hype by~\citet{tucker2022artifice}. 
}
\label{tab:intervention}
\scriptsize
\begin{tabular}{@{}p{0.35\columnwidth} p{0.60\columnwidth}@{}}
\toprule
\textbf{Instead of saying...} & \textbf{...we might more accurately say}\\ \midrule
“employers are using AI to analyze workers’ emotions” & “employers are using software advertised as having the ability to label workers’ emotions based on images of them from photographs and video. We don’t know how the labeling process works because the companies that sell these products claim that information as a trade secret.” \\ 
\midrule
"machine learning" which sounds like a computer doing something on its own & "machine training"\\
\midrule
“states use AI to verify the identities of people applying for unemployment benefits” & “states are contracting with a company called ID.me, which uses Amazon Rekognition, a face matching algorithm, to verify the identities of people applying for unemployment benefits.'' \\ 
\bottomrule
\end{tabular}
\end{table}

All in all, guarding against the fallacy of mistaking the map for the territory, we wish to strictly distinguish between humans (the kinds of beings that we are) from AI machines (knowledge models, that are at the same time currently implemented physically as artifacts, and which have real-world effects within present day human practices). At this point it is \textit{crucial} to emphasize that by differentiating humans from machines this way, we are \textit{not} defining what a human is, under an objective, materialist and reducationist paradigm, and then claiming that machines just aren’t cutting it as mechanical equivalents of whatever we defined humans to be. We certainly don’t think that currently any actual machine is cutting it,  but the more important point to make is that we claim that such a comparison is just not what matters for robot rights.   We are pointing out the pervasive misunderstanding that equates the map (machines and models) with the territory (human being) underlying robot rights advocacy.

\section{Embodied Enactive (Post-Cartesian) Perspectives on Cognition}
\label{sect:embodied}
The previous section concludes that the tools we create to understand human being, such as computational models, should not be mistaken for human being itself, a mistake rooted in implicitly already equating human being with the machinistic nature of our models and then comparing the two on their behavior. What then, do we make of our own human being, if we refuse to define it in a machinistic way? Our core philosophical starting points are embodied phenomenology and enactivism, which leads to a conception of human being as fundamentally situated and relational, governed by complex, open-ended, self-organising interactive processes. Such processes are not immaterial, but they can at the same time never be reduced to a physical state of affairs that can be engineered and built by design. On the embodied, enactive view, humans are first and foremost living beings. Human being is continuously in the process of sustaining itself, by actively engaging with its environment in a  manner that makes sense. Intelligence is sense-making: we actively relate to the world around us, and try to form a grip on the world through our actions,  as grounded in the self-sustaining of life itself. Through the process of self-individuation, living bodies continually create a distinction between themselves and their environment where none existed before they appeared and none will remain after they are gone.   Living bodies are not static organisms that passively wait to be influenced by their environments like much of the cognitivist tradition envisages.  Living bodies are fluid, networked and relational beings with values, concerns, shifting moods and sensitivities. “Living bodies have more in common with hurricanes than with statues.”~\cite{di2018linguistic}. This is true at the biological level of analysis, but extends into psychological, intersubjective, and sociocultural levels of complexity.  Living bodies laugh, bite, eat, gesture, breathe, give birth and feel pain, anger, frustration and happiness. They are racialized, stylized, politicized and come with varying ableness, positionalities, privilege, and power. We face challenges, tensions,  problems, and opportunities that constrain or enable our actions and behaviours. We are not fixed, closed systems. We are processes with a certain kind of autonomy, but this emergent autonomy is itself always in flux and has no clear boundaries. “Sense making is always under way.”~\cite[P.219]{di2018linguistic}. This provides a basis for human rights: every human being has the right to be and should be able to sustain its own autonomy in relation to its environment. From that follow questions about how one person’s right to be may interfere with another person’s rights, and how to navigate the various associated dilemmas. Outlining the precarious, socially embedded and situatedness of human being,~\citet{weizenbaum1976computer} wrote “no other organism, and certainly no computer, can  be made to confront genuine human problems in human terms. People, as complex adaptive systems, are open- ended, historical and embedded in dynamic and non-linear interactions~\cite{juarrero2000dynamics}. There is no single perfect or accurate representation or model of cognition, intelligence or emotion as such a model would require capturing cognition, intelligence, or emotions in their entirety. The circular metaphorical trick mentioned in the previous section reduces our complex, relational, intersubjective, embodied, non-determinable, non-totalizable, fluid, active and dynamic being into a set of features or a process that could be captured in data and implemented by a physical machine. But a machine is an artifact that can be achieved once a given process is complete, that is, it  is fully defined, such that it can be implemented its being rests on its closed and complete definition, whereas human being rests  on how we enact a fluid, temporarily stable autonomy, while at the same time never being fully specifiable, and always open-ended~\cite{birhane2021impossibility}.  What is interesting about actual machines, is how they mediate human being and autonomy, given that technologies can be incorporated into our interactive, embodied self-organising processes~\cite{clark2008supersizing,merleau1962phenomenology}. Machines might be used by people to empower themselves, or they might be used to oppress others. What is much less interesting, and actually distracting from the relevant ethical discussions on AI, is whether machines would become so much like the kinds beings that we are that they should have rights.

\section{
Posthumanism}
\label{sect:posthumanism}
If human being is characterized as a deeply situated and complex interactive phenomenon, and if human beings functionally incorporate parts of the (technological) environment into the flux of their self-sustaining interactions, this may be seen as an argument in line with posthumanism. In discussing the implications of posthumanism for our argument there are two versions of ‘posthumanism’ that should be clearly distinguished. We designate \textit{posthumanism}, without a hyphen, as a philosophical position that draws on thinkers like Bruno Latour to enrich understanding of the ontological status of humans and non-humans. \textit{Post-humanism} or transhumanism, on the other hand, argues that biological human being is insufficient to solve the problems in the world. Therefore, ‘we’ should (or inevitably will be) be transcended by our own intelligent technologies, for example, robots that outsmart us and outperform us. This accelerationist, transhumanist, or anti-humanist visions see human beings as irrelevant or archaic~\cite{bostrom2008want,torres2021against}. We discuss this line of thought first. 

Post-humanism or transhumanism is all too often at the center  of evolutionarily-focused robot rights arguments.  
For example, Nick Bostrom, a leading proponent of teleological posthumanism, has argued that becoming post-human – a being that transcends human capacities through, for example, augmented cognitive capacities – ought to be a central  objective for a techno-utopian future, that should be prioritized over saving human lives today~\cite{bostrom2008want,torres2021against}. Post-humanism overlaps with other similar ideologies such as transhumanism, longtermism, and eugenics, promoting a techno-utopian future where human frailty, illness, and other limitations might be eradicated   through technological enhancement, echoing the goals of the eugenics movements of the past~\cite{bostrom2008want,gebru2023eugenics}. Some transhumanists believe that it is not just human arrogance,  but  humans  as  such,  that condemn the planet to ecological crisis.  On  this view, humans, by their very nature, are simply too selfish, short-sighted, or fragile to address crises like climate change, or endure the rigors of, say, interplanetary travel. And so the future will result in artificial replacement of biological humans by something ‘better,’ beyond the merely human. If the mind could be downloaded into a metal pellet, for example, and then run  on robotics, that would be a much more resilient form of “embodiment” for these post-humanists. And if it could be done,  perhaps such pellet-copies of persons would deserve some “rights” against destruction,  as anticipated in the cyberpunk novel \textit{Altered Carbon}, in a memory-based version of historical preservation law. This type of reasoning we believe is flawed in many ways. It displays a misunderstanding of the actual nature of memory, mind, and embodiment as well as the actual natural resources (time, energy) needed to sustain it, as already discussed in the previous sections. 

While presented as post – human, transhumanism in actuality displays the arrogance of Baconian humanism taken to the next level. The fantasy of the technological copy of the human is used in one sweep to first take away attention from the actual cause of ecological crisis: the devastating effects of mechanistically techno-capitalist expansion, and then doubling down by envisioning a futuristic technological Golem that will solve our problems. Whereas this superhuman is, of course, only a metaphor for the spiraling techno-capital that was causing all the trouble in the first place~\cite{chiang2017silicon,shivani2017oculus}. The \textit{image} of the superhuman serves as a distraction from the actual challenges for human flourishing, largely caused by aggressive techno-capitalism. An \textit{actual} human transformation towards a better future may involve new technologies, however not as a \textit{replacement} (overcoming) of humans but rather as part of a more sustainable, equitable, and ecologically sound \textit{human} practices.

It should be clear that we do not believe that post-humanism as transhumanism is the answer. If anything, it represents a way of thinking that is part of the problem. A view that might provide a more thoughtful perspective is a posthumanism that calls into question the arrogance of arrogantly Baconian humanism and promotes more harmonious, just, and respectful relations between humans and the world at large~\footnote{It should be noted that~\cite{gellers2021rights} has suggested a bridge between the posthumanism-as-human-obsolescence and posthumanism- as-environmentalism positions, in the introduction of his \textit{Rights  for  Robots.}  After relating a number of ways in which robots and AI either are replacing or could replace humans, he observes that, “The two trends—the development of machines made to look and act increasingly like humans, and the movement to recognize the legal rights of nonhuman ‘natural’ entities—along with the two existential crises—the increasing presence of robots in work and social arenas, and the consequences of climate change and acknowledgment of humanity’s role in altering the ‘natural’ environment— lead us to revisit the question that is the focus of this book: under what conditions might robots be eligible for rights?”~\cite[P.4]{gellers2021rights}. Gellers’ inclusion of scare quotes around “natural” appears to adumbrate a larger goal of his project:  erosion of the categories of given and made, or natural and artificial, in order to promote the granting of the type of rights or respect offered to the former, to the latter. Unfortunately, that is a self-undermining aspect of his project; one cannot coherently use the potential or actual rights of nature as a rationale for robot rights, while simultaneously questioning the distinctiveness, utility, and  validity of the category of nature.}. Very much in line with our embodied, enactive perspective, such ecological forms of posthumanism advocate a modest, respectful self-image, in which human beings do not stand outside, over or against of the world, and the basic move of human action is not to dominate our environment, let alone ‘submit nature to our will’, as Enlightenment thinking would have it~\cite{Rushkoff2022}. On the contrary, our actions should be respectful of the fact that we can only ‘be’ because we are always already deeply embedded in complex ecological networks, and if we wish to sustain our own being, we have to be respectful and considerate to the other elements in those networks, and sensitive to the intricate interdepencies between them~\cite{bateson-2023}. On this view, moral, ethical, and political theory must move beyond considering the earth as a mere “standing reserve”~\cite{heidegger1954question} to fulfill human needs, or rather, we should acknowledge the fact that other being’s  needs are implicitly integrated in our own: my needs, are intertwined with your needs, because I depend on you and vice versa. When translated into rights discourse, this would for example balance the rights of hunters to food against the rights of animals to avoid being eaten; or the rights of off-road-vehicle users to motor through a national park against the rights of the rocks, soil, plants, and animals of the park to avoid being run over. Along these lines, law has been formed giving rights to rivers, mountains, islands~\cite{laszewska2023environmental}. 

Some robot rights advocates however lean   on this very type of ecologically informed posthumanism to argue for robot rights. They include robot technologies into the list of ‘non-humans’ that we, humans, are interconnected with, and should be considerate. The argument goes: if animals, rivers and trees are “non-humans,” that deserve rights, why not robots? Should we not avoid biological chauvinism, privileging carbon-based entities over silicon, plastic, and concrete? 

This line of argument, however, does not hold.  One cannot on   the one hand embrace posthuman language about ‘non-human’ agency, and then at the same time bring in, via  the backdoor, a quite anti-Latourian, Cartesian image of a human mind (a person with emotions, experiences, a lifeworld, a history, etc, in the everyday sense that we understand other humans), project that image onto a robot  (as in typical science fiction and in many hypothetical scenarios adduced by robot rights advocates), and then discuss that ‘artificial person’s’ rights on the basis of sentiments that are at the core quite humanist.

We can go a long way with posthuman thinking if it seems to be directed at repositioning humans in the greater context of planet earth, that is, recognizing our predicament as one of many lifeforms dependent on subtle equilibria in complex ecosystems~\cite{connolly2013fragility}. From that humble perspective  it may indeed be viable to create laws that \textit{protect }animals, rivers and trees against the oppressive tendencies of humans, because, in the end, even if only because our short-term egocentric tendencies and capitalist greed may eventually kill the very ecosystem that sustains us. 

To acknowledge the intricate, networked, and interdependent relations between human life and the environing world, consisting of all other life and the planet as a whole, does not lead to an argument in favor of robot rights, on the contrary. Robot rights proponents cannot have it both ways: \textit{either} robots deserve rights because they are special as opposed to other forms of matter, such as rocks, or tables, or bits of raw material. \textit{Or}, on the basis of a networked posthuman ecological vision, robots are not special, but simply one element mediating in larger interconnected eco-systems in which living and non-living things hang together and sustain each other. In the first case, robot rights defenders implicitly return to a \textit{humanist} argument: robots deserve rights, because of some special characteristics that robots have, or are about to have in the near future. But in that case the argument cannot rest on a posthuman worldview. And the second part of the argument can in that case not be that robots are ‘non-humans’ that should have rights, where in a humanist paradigm they were denied so. In this case, robot right defenders still treat robots implicitly as sentient beings ‘like’ humans, and then wish to grant them rights on this, at the core humanist, basis. 

Alternatively, one fully endorses posthuman thinking, which means rejecting outright all classic modern, humanist distinctions between subject and object, and between materialist science and social-political theory, as Latour advocates~\cite{latour-1998}. But before we start to discuss the moral status of the robot within such a radical posthuman paradigm, we should first acknowledge that the robot, as we have conceived, designed and engineered it so far, is the very anti-thesis of this movement. Up until now, the robot has been  the paragon  of human technological domination over the earth, based on detached human cognition that puts ourselves outside of ‘nature’ and tries to control it. In the posthuman image, the robot, \textit{as we have both understood and designed it so far}, is not ‘another non-human’ \textit{like} animals, rivers and trees, that we should acknowledge and perhaps grant rights, courtesy of being not just means but ‘ends in their own right’~\cite{latour-1998}~\footnote{~\cite{latour-1998,latour2007modernize} discusses quasi-objects and non-humans where it is clear that by \textit{non-human} he does not mean ‘nature’ (as opposed to human) with this, he rather wishes to transcend the dichotomy. Latour outlines that \textit{all} things, not just humans, are never just ‘means’ but also ‘ends in and of themselves’ (in reference to Kant). For example, for Latour, the river is an end in itself in the sense that the river ‘has its own way’. One might try to make a straight functional canal out of it, but engineers discover that the river will then create erosion or flooding or other problems because its way has been denied. We do not object any of this. However, it is important to emphasise that Latour does not make a distinction between ‘human-made technologies’ and  other things. Latour, for example, does not discuss “the way of the canal”. Is the canal, once it is created by human engineers’, also a thing that has an ‘end’ in and of itself, just like the river? We contend it does not. How can it be as it was simply used as the example of human engineering \textit{denying} the river its own being? Should we now discuss the river and the canal and whether we should give both of them ‘their space’? This is what robot rights amounts to when one follows this line of argument. Robots are \textit{not} rivers, they are canals. And while we might discuss our intricate ecological relations as humans to rivers, it is problematic to treat canals and robots in the same way, precisely because canals and robots represent the modernist perspective that Latour is arguing against.}. The traditional robot \textit{is} only a means. It \textit{denies} the idea of a posthuman sphere in which non-humans have intrinsic value. It is therefore the thing we need to strictly limit in order for us to break free from traditional humanism and dominionism, and in search of a better way to relate to the planet at large~\cite{white1967historical}. The robot, so far, is thus the physical representative – the ultimate model – of anti-posthuman, modernist thinking. So if we indeed wish to fully embrace posthumanism in a Latourian sense, we need to get rid of ‘the robot’ first. If we embrace posthumanism, ‘the robot’ that we were deliberating, will disappear altogether, and human being in its traditional humanist image will dissolve along with it as well, because the two go together. What will emerge instead are complex constellations of interconnected bio-socio-technical dependencies, which cannot be understood, and therefore not in totality engineered in any finite sense~\cite{latour-1998}. In such a situation, the question of rights is an open question, which needs nuanced discussions centered around concrete real-world projects, instead of purely theoretical argumentation. The vocabulary used by Latour can at times be unhelpful in grasping his alternative worldview. , In talking about ‘humans’ and ‘non-humans’ as a collection of ‘things’ making up these ‘networks’, a. potential danger arises where the reader might interpret these object-words once more as objects. Whereas the idea of an ‘object’ is itself a modern construct (it immediately implies the existence of a \textit{subject} detached from it~\cite{husserl1922ideen,heidegger1962being,merleau1962phenomenology}. 

While posthuman thinking may help towards developing new practices for relating to the world, and for understanding ourselves as human being enmeshed in the world, in no way does it imply robot rights. At best, robots might be included in discussion about rights that have to do with constellations and relations, with systemic interdependencies. Thus~\cite{coeckelbergh2010robot} states: “we can arrive at a range of forms of moral consideration that will be less strong than robot rights but still imply some obligations towards robots in the context of particular human-robot relations.”~\cite[P.218]{coeckelbergh2010robot}. But that characterization obscures more than it reveals. The moral obligations towards robots, in any practical scenario we would consider meaningful, do not concern the \textit{robots}, but rather the distributed networks of human-environment relations. For example, if a person is dependent for sustaining their own being on robot assistance, it might be a case of moral injustice to damage that robot. But this moral consideration concerns preserving the integrity of the distributed networked set of relations in which the robot is one element. The largest such network would concern the integrity of the planet at large. But this has been our conception of \textit{human being} all along: human being \textit{is} a deeply situated and distributed relational activity, which sustains itself, in action (see section~\ref{sect:embodied}). In this perspective \textit{all} things, also things that are neither traditionally human nor robot, would \textit{be} eligible to the very same moral considerations. One might be dependent on a robot, but also on a prosthetic leg, a wheel-chair, a pacemaker, an ergonomic table, clothes, a writing book, glasses, or a fountain pen. In the posthuman view, there is \textit{nothing} special about the \textit{robotness} of robots that would form an argument for robot rights. Instead, posthumanism, in our view, mostly \textit{extends} current debates on \textit{human} rights, acknowledging that our human being is dependent on complex ecological networks of which we are part. 

More concretely, robot rights proponents have yet to grapple with the enormous environmental damage caused by the current robots and AI that are designed on the basis of humanist and modern foundations. AI is not just abstract algorithms but includes the wider social, political, cultural and environmental infrastructure. As~\cite{crawford2021atlas} argues, from personal assistants such as Siri to the mobile devices and laptops to “self-driving” cars, AI cannot function without minerals, elements, and materials that are extracted from the earth. “AI system, from network routers to batteries to data centers, is built using elements that required billions of years to form inside the earth.”~\cite[P.31]{crawford2021atlas}. Crawford lists 17 rare earth elements including lanthanum, cerium and praseodymium, which are making our smart devices smaller, lighter and more efficient. But extracting, processing, mixing, smelting, and transporting these elements and minerals imposes massive costs, including environmental destruction, local and geopolitical violence, and human suffering. For example, half  of the planet’s total consumption of lithium-ion, a crucial element for Tesla Model S electric car battery, is consumed by Tesla~\footnote{“[O]nly 0.2 percent of the mined clay contains the valuable rare earth elements. This means that 99.8 percent of earth removed in rare earth mining is discarded as waste, called ‘tailings,’ that are dumped back into the hills and streams, creating new pollutants like ammonium”~\cite{crawford2021atlas}.}). Furthermore, between 2010 and 2018 alone, computing workload required for data centers has increased nearly by 550\%~\cite{siddik2021environmental}. In 2014, a total of 626 billion litres of water was used by US data centres for electricity generation and cooling~\cite{mytton2021data}. 

Thus the advocacy for robot rights presented as a natural outgrowth of environmentalist thought is jarring, given that, “[i]f we visit the primary sites of mineral extraction for computational systems, we find the repressed stories of acid-bleached rivers and deracinated landscapes and the extinction of plant and animal species that were once vital  to the local ecology”~\cite[P.36]{crawford2021atlas}. The creation of AI then can be felt in the “atmosphere, the oceans, the earth’s crust, the deep time of the planet.”~\cite{crawford2021atlas}. Although much of the energy consumption by AI models is a closely guarded secret,~\cite{belkhir2018assessing} estimate the tech sector will contribute 14 percent of global greenhouse emissions by 2040 (see also,~\cite{Brevini21,dobbe2019ai}).~\cite{bender2021dangers} further highlight how incredibly resource intensive AI systems are: “Training a single BERT base model (without hyperparameter tuning) on GPUs was estimated to require as much energy as a trans-American flight”~\cite{bender2021dangers}. Similarly,~\cite{dodge2022measuring}'s found that fine tuning a partially trained large language model with 6 billion parameters, for example, “emits more CO2 than all emissions from an average US home for a year (which includes emissions from electricity generation, natural gas, liquid petroleum gas, and fuel oil, totaling 8.3 metric tons CO2 per year)”. Given the increasing material and environmental cost affecting the climate and ecosystem, we argue putting AI in the same category as the environment under the banner of ‘non-human’ entities, is a grave category error, akin to putting Donald Trump in the category of civil libertarian freedom fighter because both fought “censorship.”

\section{The Legal Perspective}
\label{sect:legal}

Many legal scholars have developed creative arguments for recognizing certain rights of robots themselves, or of the persons and corporations who manufacture, own, and control them. At least two broad categories of rights are at issue. First is a right of physical integrity–the sense that a robot, like a recognized work of art protected by the ”moral rights” (a copyright law term of art) recognized under the Berne Convention, should be preserved from destruction,  or even from damage.  This “freedom from” damage,  rather than “freedom to” act,  is the easiest to defend,  but is still unconvincing when framed as the right of the robot, rather than the property right of its owner or controller. As~\cite{schroder2021robots}, rules regarding the treatment of robots are “ultimately for the sake of the humans, not for the sake of the robots”~\cite[P.200]{schroder2021robots} citing~\cite{darling2013giving}.
 
More problematic are alleged rights of robots to act—for example, to speak, marry, own property, vote, or contest discrimination. This section critically examines such alleged positive rights of robots, focusing specifically on legal arguments and arrangements. 

Consider first the potential rights of a robot to marry or vote. We might fruitfully dissect such a right by considering a comparison with the realm of art mentioned above. Consider the regard a quilt collector may have for a particularly intricate and storied blanket, or a museum director may have for a painting by Remedios Varo or George Grosz. Few would object to either person devoting resources to preserving such artifacts from harm, writing extensively about their nature and meaning, and enthusiastically showing them off to friends and visitors. Yet there would be a justifiable concern if either started talking to the quilt or painting, proposed marriage to it, or insisted it be permitted to vote—even if a clever designer placed a small microphone and speaker behind each with access to a chatbot which permitted it to appear to insist upon its own sentience. The utter dependence of such an absurd apparatus on sociotechnical (and, at present, powerful corporate) support belies its simulation of animation. Moreover, given the opacity of AI training, we have no idea what agenda the corporately animated artifact would have, and how far it could deviate from the goals of Microsoft, Google, or whatever other firm ultimately controls it. 

This may seem like a bizarre thought experiment, but far more unrealistic examples appear in the robot rights literature. Consider this example from one of the leading thinkers addressing the issue affirmatively,~\cite{danaher2020welcoming}. Danaher offered the following as one of many defenses of his position of “ethical behaviorism:”

\begin{quote}
    “[S]uppose someone woke up one morning and was told that their spouse of the past 20 years is an alien from the Andromeda galaxy. The doctors have performed tests and it turns out that they have an entirely silicon-based biology. Nevertheless, they still act in the same way, behave in the same way, and appear to be the same loving and supportive spouse that they always were (albeit with some explaining to do). Would the knowledge that they are made of different stuff warrant their being denied the moral status they have always been afforded? Or would the behavioural evidence negate the relevance of this new bit of information? The latter is the more plausible view than the former: it would require remarkable cruelty and indifference to their day-to-day interactions to reject their moral status.”
    \null\hfill~\cite[P.2032]{danaher2020welcoming}
\end{quote}

While philosophy and law should remain open to insights from the humanities, they must be more thoughtful than such a contrived hypothetical narrative. All the love and support postulates by Danaher may merely have been simulated, and thus false and deceptive, underscoring a deep flaw of behaviorist theories: intent matters. To begin to entertain the plausibility of this thought experiment, the reader must conjure a scenario where such life forms had been discovered, and some critical mass of them had proven to be exceptionally kind-hearted (or merely innocuous). But why do so? If science fiction is fair game for philosophical reasoning, why not also contemplate the possibility that the Andromedans are playing a “long game” and will take over the planet once a critical mass of them has infiltrated earth institutions? Indeed, after publication of Cixin ~\cite{liu2015dark}’s \textit{The Dark Forest}, one staple consideration for thinkers at the intersection of science fiction and philosophy must be a chillingly plausible “solution” to Fermi’s Paradox: that the main reason we on earth have not discovered extraterrestrial life is because any civilization sufficiently advanced to be discovered would be tempted to destroy any less advanced civilization that contacted it, in order to avoid any threats that the rising civilization could pose to it. 

Other legal discourse on robot rights discusses such hypothetical scenarios with more evenhandedness. For example, while Simon Chesterman acknowledges that “arguments about ‘rights for robots’ are presently confined to the fringes of the discourse,” he postulates that “AI systems truly indistinguishable from humans might one day claim the same status” and rights as human beings”~\cite[P.116]{chesterman2021we}. Moreover, he warns that, “Though general AI remains science fiction for the present, it invites consideration as to whether legal status could shape or constrain behaviour if or when humanity is surpassed. Should it ever come to that, of course, the question might not be whether we recognize the rights of a general AI but whether it recognizes ours”~\cite[P.116]{chesterman2021we}. Yet if the possibility of such a “robot take-over” is even thinkable, that would seem to be a very strong argument in favor of the strict restrictions upon robots, rather than any grant of rights to them. 

Addressing more quotidian scenarios, legal scholars~\cite{froomkin2015self} cleverly derive a reasonable expectation of robot owners that their machines are to be left unmolested, by considering the proper scope and limits of self-defense against robots that could either harm or spy on humans. Responding to reports of a property owner who shot down a drone overflying his property, as well as many other incidents and possibilities, they conclude that “the best way to create a balance–to ensure that drones are not unduly attacked and that people are only duly worried about what drones are doing–is  to standardize how drones–as well as other robots–declare their capabilities and intentions, thus changing what fears   are legally and morally reasonable.”~\cite{froomkin2015self} focus on the property rights of the robot’s owner, for, as they accurately note, “in law, at present, a robot has the same rights as a sock.” The owner has no right to expect others to freely permit a robot to invade their property and personal space. However, it is not hard to extrapolate from ~\cite{froomkin2015self}’s proposal a reciprocal duty to respect others’ property that should bind those who encounter robots whose owners have credibly given notice are not threatening.~\cite{lemley2019remedies} take a similarly measured approach, suggesting that owners of robots have a basic due process right to be properly compensated before government damages or destroys them. But even this approach must be cabined, like free speech rights, with a time, place, and manner exception: no government should be expected to go to great lengths to compensate a robot owner whose creation (or swarm of creations) creates public nuisance. And most importantly for our purposes, all the modest rights mentioned so far accrue to a robot’s owner, not to the robot itself.

A second strand of robot and AI rights arguments in law, focusing on speech, tends to be more expansive than  the literature on a presumption of continued existence absent some good reason for dismantling.~\cite{massaro2016siri} have advanced one of the most legally sophisticated cases for this position. In the article “Siri-ously?,” they put forward an audience-focused theory of freedom of expression. Rather than focusing on the autonomy or flourishing of  speakers, such a theory instead concentrates on the rights of listeners. From this perspective, the key actors in public discourse are those that “produce information useful to natural persons who seek to participate in public discourse. That a computer, not a human, produces the useful information should not matter.”~\cite[P.1160]{massaro2016siri}. Nevertheless, they are still focused on the human in their argument. As they conclude, “because a primary basis for protecting speech rests on the value of expression specifically to human listeners, free speech protection for strong AIs that attends to the value (and dangers) of such speech to such listeners does not rob the First Amendment of a human focus.”~\cite{massaro2016siri}. In other words: just as the US First Amendment might invalidate laws banning pens, because of the expression they enable, so too might it invalidate a ban on sound-emitting robots. But to call this invalidation a recognition of the right of the robot would be as implausible as identifying “pen rights” in the case of an invalidated pen ban. The robot is a tool to help ensure the realization of persons’ rights, not an independent rights-bearing entity.

Nevertheless, in a Google-sponsored white paper,~\cite{volokh2011google} make an even more expansive argument in the course of arguing for free expression rights for firms that own and operate search engines. They tend to  focus on speech in the abstract as the object to be respected and maximized, rather than any particular entity’s  rights: “First, Internet speech is fully constitutionally protected. Second, choices about how to select and arrange the material in one’s speech product are likewise fully protected. Third, this full protection remains when the choices are implemented with the help of computerized algorithms. Fourth, facts and opinions embodied in search results are fully protected whether they are on nonpolitical subjects or political ones. Fifth, interactive media are fully protected. Sixth, the aggregation of links to material authored by others is fully protected. Seventh, none of this constitutional protection is lost on the theory that search engine output is somehow “functional” and thus not sufficiently expressive.  And, eighth, Google has never waived its rights to choose how to select and arrange its material”~\cite[P.7]{volokh2011google}. 

Given its immediacy, the right to “speak” was one of the earliest claims for rights made on behalf of those operating  or creating AI, robots, and automated entities~\cite{benjamin161algorithms}. Google has repeatedly asserted that search results (almost entirely algorithmically generated) should be protected by First Amendment law~\cite{bracha2008federal}. Lower courts have recognized that right, and many cyberlibertarians hail this outcome as a triumph—forbidding, for example, an authoritarian administration from forcing the firm to disappear results leading to its critics. However,  as a practical matter, the seminal case(s) where Google asserted this right involved business torts, where a company asserted that it had been ranked unfairly. Had it been adopted in Europe, such a logic easily could have scuttled the European Commission’s successful antitrust case against Google, since the remedies in the case require the company to present information in ways it does not wish to communicate.

But note once more, to return to the fundamental point: the search algorithm or servers implementing it were not  the rights-bearing entities in any of the Google cases. Rather, the corporation Google itself was. Whatever one’s views on the propriety of corporate speech  protections, the bottom line here is that it is difficult to frame a coherent case for any robot or AI itself   to have rights.

\section{The Troubling Implications of Legal Rationales for Robot Rights}
\label{sect:implications}

Legal discourse about rights is distinctive from ethical, metaphysical, and philosophical discourse about rights, though it also informs (and is informed by) ethics and philosophy. For a pragmatic attorney, rights talk \textit{must} “cash out”  in some actionable claim in courts of law, or before administrative tribunals.  Envisioning how such rights claims would actually redistribute resources and respect in society quickly demonstrates some disturbing implications of  even the most anodyne proposal for robot rights.  For example,  several advocates for  robot rights  claim that  they are expanding the circle of human concern, from persons, to features of the natural environment, to features of the built environment.  However, even the prospect of heightened concern for (let alone rights for) a robotic police force is extremely disturbing to many advocates of hyper-surveilled minoritized communities. Consider, for instance, U.S. Congressman Jamaal Bowman’s response to a police robot inflicted upon New Yorkers by a police department that has been widely condemned as having engaged in racist practices~\cite{harcourt2018counterrevolution}.

\begin{quote}
  We scream defund the police so we can allocate those resources towards something that focuses on true public health and public safety. Protesting all summer, for Black lives, we were under assault. People living in poverty, struggling, struggling to put food on the table, keep a roof over their head, take care of their kids, afford child care. All this going on. And now we got damn robot police dogs walking down the street. What the hell do we need robot police dogs? This is some RoboCop shit, this is crazy. (~\cite{bowman21} in Newsweek).  
\end{quote}

Now imagine one of the advocates for a robot right not to be “harmed”, destroyed, or turned off, trying to  reason with Bowman. “Don’t you see, if we just respect this robotic police dog, we’ll eventually grow in moral concern  for all?” Or, worse: “Do not feel you have any entitlement to break a leg off the policing robot if it subdues you, even if you are innocent. That would be akin to animal abuse.” These types of responses would be condescending at best, and cruelly indifferent to the real social realities of robot rights assertions at worst. Yet they are the potential implications of many of the ethical and philosophical commitments to robot rights explored in the sections above. 

More troubling outcomes are likely if advocates for humanoid robots claim human rights for them or program them to imitate a desire for such rights  and  convince  authorities  to  respect  them.  Consider,  for  instance,  the  right to communicate online – lawyer~ John Frank~\citet{weaver18} has argued that “all the constitutional speech protections that humans enjoy in the United States” should “apply to A.I., robots, and Twitterbots, whether they are Russian intruders or Microsoft mistakes.” The proliferation of bots and output from generative AI has already overwhelmed and drowned out actual human speech in several contexts, spamming hashtags, amplifying stereotypes, and otherwise impeding communication and collective will formation~\cite{bianchi2022easily}.  The larger purposes of free expression are to enable human autonomy, and to allow humans to democratically self-govern. These purposes are not served by expansions of free expression rights to bots (often built by already-powerful entities) that would permit them to dominate communicative spheres.

Matters deteriorate even further if we consider possible applications of free expression rights for, say, insect-sized  drones that can project images, films, or speeches to individuals. The right of humans to peaceful enjoyment of public spaces may be gone forever, as persons struggle to fight off a near-constant ”communication” of advertisements and other messages.  Nor should advocates of robot rights forget the rather robust rights to record that many humans have been granted by courts~\cite{glik2011}. These rights have been granted in the context of human limitations—the common sense presupposition that few, if any, persons will be creepy and persistent enough to stalk others with a cell phone camera. Attach the same camera to a drone, however, and all bets are off. Privacy may be  eviscerated~\cite{nourbakhsh2013robot,veliz2020privacy}.

But even if the interests involved were closer to equipoise, it’s crucial to ask: what does the attribution of constitutional rights to robots add to the discussion? What is the “interest” of the robot here? If it has been programmed to “want” to maximize the number of persons who see its ads, perhaps it “feels” some “pain,” or whatever a robotic analogue of discomfort or disappointment would be. But it could be reprogrammed for a different optimization function~\cite{pasquale2020new}. As~\citet{mosakas2021moral} argues in “On the Moral Status of Social Robots: Considering the Consciousness Criterion,” there is no evidence in robotics of anything like the fixed consciousness of animals and humans that demands moral concern. This is a conclusive argument against a position, like Jowitt’s, that basing legal rights on human embodiment is arbitrary~\cite{jowitt2021assessing}. Indeed, the most likely outcome of a principle of non-discrimination between humans and machines would be to further tip the scales of societal influence toward the large firms now best poised both to deploy massive numbers of robots and AI, and to mobilize them to better manipulate judicial, political, and communication systems to their advantage.

In the abstract, the contemplation of robot rights seems extraordinarily cosmopolitan and open-minded—part of  the progress of history toward including more entities as entitled to moral concern. However, as with debates over the expansion of social rights, there must be an acknowledgment of the costs of such rights, including the responsibilities they may impose on natural persons while absolving powerful bodies behind robots and AI of responsibility and accountability.

Whatever the science fictional appeal of speaking robots, no one should romanticize manipulative bot-speech. The logical endpoint of laissez-faire in an automated public sphere~\cite{pasquale2017automated} is a continual battle for mindshare by various robot armies, with the likely winner being the firms with the most funds, power, and influence. They will micro-target populations with “whatever works” to mobilize them (be it truths,  half-truths,  or manipulative lies),  fragmenting the public sphere into millions of personalized, private ones. Such an endpoint  does not  represent  a triumph  of classic values  of free expression (autonomy and democratic self-rule); indeed, it portends their evaporation into the manufactured consent of a phantom public~\cite{herman2010manufacturing,lippmann1925phantom}. Undisclosed bots counterfeit the basic reputational currency that people earn by virtue of their embodiment~\cite{pasquale2020new}.

\citet{gunkel13} has tried to refocus the robot rights debate from whether machines can acquire human-like capabilities,  to a concern about how “anthropocentric criteria” may “not only marginalize[] machines but [have] often been mobilized to exclude others—including women, people of color, and animals.”~\cite{gunkel13}. While the idea    to abandon the human as a locus of exclusive concern and to recognize past exclusions of persons as grave injustices, this should not be trivialized by comparing them with the putative exclusion of robots from the relevant circle of concern and respect.

To put this more concretely, consider worldwide conflicts over same-sex marriage rights. Promoters of robot rights may argue that just as a person’s ability to marry a longtime partner, regardless of gender, is an emancipatory project, so too is the ability of a person to marry a robot. But these are not similar ambitions, and to compare the one with the other is to trivialize the larger project of equality for sexual orientation and minoritized gender identities. Indeed, the religious right has in the past used exactly the type of flattening or relativistic rhetoric characteristic of many robot rights supporters to dismiss the cause of gay marriage, likening it to a person trying to marry their dog   or “box turtle,” in the strange formulation of one US Senator~\cite{dallasvoice2010}. The line between post-human and dehumanizing rhetoric can be thin, and the former is easily repurposed toward the latter ends.

What about a right not to be harmed? MIT robot ethicist Kate Darling has proposed that “treating robots more  like animals could help discourage human behavior that would be harmful in other contexts”~\cite{darling2016extending}. For example, kicking a robotic turtle may be an early sign of sociopathy. However, this line of argument falls under an argument for the well-being of the human involved in the robot-human relationship that we have covered in Section~\ref{sect:posthumanism}. Rules against it would perhaps lead to more pro-social behavior. However, complicating the hierarchy of machine, animal, and human could also lead to some inhumane outcomes. In the U.S., an animal shelter purchased a K5 Knightscope robot to patrol its grounds, in part to keep homeless persons away. Rebellious would-be campers retaliated by tying  a tarp around the robot, covering its sensors with barbecue sauce, and knocking it over. A shelter spokeswoman lamented how much her employer, a non-profit, had to spend on security. That lament was greeted with a tide of derision online. The shelter had apparently failed to consider whether a relentless robotic gaze on its grounds might  be cruel to persons.

But what if the robot were deemed to deserve something like animal rights, or even animal welfare protections? The rebellious campers might be charged with animal cruelty and face jail time. However, the types of rights and interests protected by laws against animal cruelty do not really exist in this situation: the robot cannot feel, and has no phenomenological engagement with the world remotely similar to that of humans and those animals protected by animal welfare laws. That is one reason why humans are permitted to walk freely on the sidewalk, whereas the same right should not be given to robots (which are better classified once again as our tools, not our partners: as vehicles and properties in need of a license).  In the US, The San Francisco Department of Public Works has warned that unauthorized operation of a robot on sidewalks risks a daily fine~\cite{Cuthbertson17}, and it is correct to do so: humans can feel real pain if run over by a robot, but the same stakes are not there for robots themselves (which cannot feel anything, and cannot be harmed by being turned off or reduced to scrap metal). All these practical problems are yet more vindications of~\citet{solum1992legal}’s prescient cautions in 1992. It would be deeply unwise to confer legal personhood on robotics and AI. As over 280 European experts have concluded, ”creating a legal status  of electronic ’person’ would be ideological and non-sensical and non-pragmatic”~\cite{nevejans2018open}.

Given the remarkable affordances of robotic technology to surveil, oppress, and marginalize persons, we should interpret via a hermeneutics of suspicion repeated efforts to build up a generalizable moral duty to respect robot rights from occasional positive human emotional responses stimulated by robots, or simulation of human relationships between humans and robots. As our response to the thought experiment regarding the “Andromedan alien” posited by Danaher above noted, the composition, origin, and intent of an entity matters critically to the entity’s reception as a moral agent or patient. One cannot simply elide such critical considerations by positing an ethics of relationality.

\section{The Enduring Irresponsibility of AI Rights Talk}
\label{sect:conclusion}

Robots rights discourse has become an active area of philosophical inquiry, despite the warnings of leading thinkers in the philosophy of technology that 
such concerns are misplaced. As~\cite{schroder2021robots} has observed, Luciano Floridi, Richard Coyne, Joanna Bryson, Alan Winfield and Margaret Boden all warned against such rights attribution, with Floridi in particular dismissing it as “distracting and irresponsible”~\cite{floridi2017robots}. Despite such considered judgments, the discourse continues on, and below we look at two related and foundational errors in reasoning behind such discourse. 

The first is a focus on relationships without due regard to realities, which often leads to meta-ethical frameworks that slide too readily between “is” and “ought.” For example, Mark Coeckelbergh pioneered a “social-relational” approach to the question “should we grant rights to robots”~\cite{coeckelbergh2010robot}. Coeckelbergh observed that “A social ecology is about relations between various entities, human and non-human, which are inter-dependent and adapt to one another. These relations are morally significant and moral consideration cannot be conceived apart from these relations.” He claims that “both defenders and opponents of robot rights” might take inspiration from his approach. Yet he also claims that, “in this approach to moral consideration it no longer makes sense to talk about moral consideration of robots in general, for example robot rights,” in part because robots are so diverse. This concession of the unhelpfulness of “robot rights” is heartening.~\cite{coeckelbergh2010robot} further says: “On this basis, we can arrive at a range of forms of moral consideration that will be less strong than robot rights but still imply some obligations towards robots in the context of particular human-robot relations.” This, we argue, points to a departure from robot rights and towards human rights. It is analogous to saying that we are not allowed to damage a persons’ wheelchair, where the concern is not about the wheelchair, it is about the person that has ‘incorporated’ the wheelchair into her being (a topic we discussed in Section~\ref{sect:posthumanism}). A social-relational approach nonetheless is in constant danger of eliding the critical consideration that persons should want to know exactly how a robot was programmed, what it is made of, what its purposes are, and how it can be controlled—in other words, \textit{what it is}—before engaging with it, let alone commencing a social relationship with it. To the extent they have failed to do so, and have begun engaging in simulations of social relations with robots without such due diligence, this is less a form of support for particular robots’ rights, than evidence of the deployment of deception and dark patterns by technology creators and vendors that deserve the burden of strict regulation, not the reward of rights.

The second error is rhetoric that treats machines as entities that can be othered, oppressed, harmed, or discriminated against, which so often fails to comprehend (or is willingly ignorant of) the fact that machines are tools. They may be helpful or destructive, well or poorly made, reflective of human values, or destructive of them. The reality remains that numerous forms of AI and robotics are first and foremost tools used by powerful entities (government agencies, powerful institutions, and big tech alike) to harm, surveil, track, monitor, oppress and discriminate against society’s most vulnerable~\cite{mcquillan2022resisting,noble2018algorithms,eubanks2018automating}. Whatever their ultimate value, they are not on a moral level with humans or even non-human animals. To think otherwise is to risk making humans the “tools of our tools,” to paraphrase Henry David Thoreau (\citeyear{thoreau-1854}).

To answer the question posed at the beginning of this article–is it time to stop asking what machines can do for  us, and to ask instead what we can do for machines–the answer is a resounding no. Indeed, the question only makes sense from a position of privilege, where one is not daily experiencing, say, the surveillance of a drone, the harassment of a security guard robot, the overwhelming military superiority of a hegemonic power, spamming by bots, and any number of other actual and potential infringements of human liberty and welfare by machines.

Despite so much injustice and harm mediated by robotic systems, we can still imagine and envision a future that is kinder, more humane, and more just, and robotic technologies can certainly play a role in such a future. Nonetheless, rights for robots, or even the idea of rights talk, is not the way to it; it is the opposite. While robot right defenders try to argue that rights for robots help marginalized groups, reflect and inculcate concern for the natural environment, and would on the whole be ethically just (assuming a future sentient robot), we have argued that the opposite is true: robot rights’ most important effects would be the advancement of corporate power (and a mechanism to evade responsibility and accountability) at the cost of the marginalized. The idea of granting rights to a future  sentient robot legitimizes a kind of techno-optimist thinking which, much like the current fad of commercial space travel, actually undermines rather than promotes sustainability~\cite{torres2023acronym}. Instead, if a just future and relationship with our technologies is the goal, what we need is legislation for accountability for the technological spheres now dominated by powerful corporations. The historical and current crises of racial subordination, economic inequality, and climate change should be the impetus for such legislation. Machines must be developed to address these crises, rather than treated as having any type of ethical or legal status themselves

\printendnotes

\bibliographystyle{apalike}  


\bibliography{references}

\clearpage 
\appendix


\end{document}